\documentclass[12pt]{article}
\usepackage{axodraw}
\input epsf.tex
\usepackage[svgnames]{xcolor}

\newcommand{\beq}{\begin{equation}}
\newcommand{\eeq}{\end{equation}}
\newcommand{\beqa}{\begin{eqnarray}}
\newcommand{\eeqa}{\end{eqnarray}}
\newcommand{\beqar}{\begin{eqnarray*}}
\newcommand{\eeqar}{\end{eqnarray*}}

\newcommand{\eps}{\epsilon}
\newcommand{\ga}{\gamma}

\newcommand{\inn}{\!\cdot\!}

\newcommand{\eg}{{\it e.g.,}\ }
\newcommand{\ie}{{\it i.e.,}\ }
\newcommand{\labell}[1]{\label{#1}} 
\newcommand{\reef}[1]{(\ref{#1})}
\newcommand\prt{\partial}
\newcommand\veps{\varepsilon}

\newcommand\cR{{\cal R}}
\newcommand\cA{{\cal A}}
\newcommand\cM{{\cal M}}

\newcommand\cJ{{\cal J}}

\newcommand\cG{{\cal G}}
\newcommand\cI{{\cal I}}

\newcommand\cP{{\cal P}}

\newcommand\bz{\bar{z}}

\newcommand\Tr{{\rm Tr}}

\begin{document}

\vspace*{1cm}

\begin{center}
{\bf \Large T-dual RR couplings on D-branes \\
 from S-matrix elements
}

\vspace*{1cm}

{Komeil Babaei Velni}\footnote{komeilvelni@gmail.com}\\
\vspace*{1cm}
{ Department of Physics, Facualty of Science, University of Guilan, Rasht, Iran}
\\
\vspace{2cm}

\end{center}

\begin{abstract}
\baselineskip=18pt

Using the linear T-dual ward identity associated with the NSNS gauge transformations, some  RR couplings on D$_p$-branes have been found at order $O(\alpha'^2)$. We examine the $C^{(p-1)}$ couplings with the S-matrix elements of one RR, one graviton and one antisymmetric B-field vertex operators. We find the consistency of T-dual S-matrix elements and explicit results of scattering string amplitude and show that the string amplitude  reproduces these couplings as well as some  other couplings. This illustration is found for $C^{(p-3)}$ couplings in the literature which is extended to the  $C^{(p-1)}$ couplings in this paper. 

\end{abstract}
\vskip 0.5 cm


\vfill
\setcounter{page}{0}
\setcounter{footnote}{0}
\newpage
\section{Introduction} \label{intro}
The dynamics of the D-branes of type II superstring theories at the lowest order in $\alpha'$ is  given by the   world-volume theory which is the  sum of   Dirac-Born-Infeld (DBI) and Chern-Simons (CS) actions  \cite{Leigh:1989jq, Bachas:1995kx, Polchinski:1995mt, Douglas:1995bn}. The bosonic part of this action is  
\beqa
S&=&-T_p\int d^{p+1}x\,e^{-\phi}\sqrt{-\det\left(G_{ab}+B_{ab}\right)}+T_{p}\int_{M^{p+1}}e^{B}C+\cdots\labell{DBI}
\eeqa
where dots refer to the   terms at higher order of $\alpha'$. In the Chern-Simos part, $M^{p+1}$ and  $C$, represents the world volume of the D$_p$-brane and the sum over all  RR potential forms, respectively. 
 The multiplication rule is  wedge product.
 The closed string fields $G_{ab}$ and $B_{ab}$ are  the pulled back of the bulk fields $G_{\mu\nu}$ and $B_{\mu\nu}$ onto the world-volume
of D-brane.   
 The action to be invariant under the B-field gauge transformation by adding the abelian gauge field as $B\rightarrow B+2\pi\alpha'f$. The Chern-Simons part is invariant under the RR gauge transformations $\delta C=d\Lambda+H\wedge \Lambda$ in which $H$ is the field strength of $B$ and $\Lambda=\sum_{n=0}^7\Lambda^{(n)}$.

The couplings of $D$-brane at order $\alpha'^2$ involving the linear  RR field strengths $F^{(p)},\, F^{(p+2)}$ and $F^{(p+4)}$ have been found in  \cite{Garousi:1996ad}. It has been shown that these couplings are invariant under the linear T-duality and the $B$-field gauge transformations.

 These couplings for $F^{(p+2)}$  in the string frame are \cite{Garousi:2010ki}:
\beqa
&\!\!\!\!\!\!\!\!\!&\frac{2\pi^2\alpha'^2T_p}{p!}\int d^{p+1}x\,\eps^{a_0\cdots a_p}\left(\frac{1}{2!} { F}^{(p+2)}_{ia_1\cdots a_pj,a}\cR^a{}_{a_0}{}^{ij}-\frac{1}{p+1}{ F}^{(p+2)}_{a_0\cdots a_pj,i}\hat{\cR}^{ij}\right)\labell{LTdual}
\eeqa
where $F^{(n)}=dC^{(n-1)}$ ,  $\cR$ is the linear curvature tensor, $\hat{\cR}_{ij}=\frac{1}{2}(\cR_{ia}{}^{a}{}_{j}-\cR_{ik}{}^{k}{}_{j})$  and commas denote partial differentiation.

These quadratic  couplings can be extended to higher order terms by making them to be covariant under the coordinate transformations and invariant under the RR gauge transformations.
 That is, the partial derivatives  in that couplings should be replaced by the covariant derivatives, the closed string tensors should be extended to the pull-back of the bulk fields   onto the world-volume of D-brane,  the linear RR gauge field strength   should be extended to the nonlinear field strength   $\tilde{F}^{(n)}$ that is given as
\beqa
\tilde{F}^{(n)}=dC^{(n-1)}+H\wedge C^{(n-3)}\labell{tF}
\eeqa
  and the linear curvature tensor should be extended to the nonlinear curvature tensor $R$, \ie  
\beqa
\!\!\!\!\!\!\!\!\!\frac{2\pi^2\alpha'^2T_p}{p!}\int d^{p+1}x\,\eps^{a_0\cdots a_p}\left(\frac{1}{2!}{\tilde{F}}^{(p+2)}_{ia_1\cdots a_pj}R_{aa_0}{}^{ij;a}+\frac{1}{p+1}{ \tilde{F}}^{(p+2)}_{a_0\cdots a_pj;i}( R_a{}^{iaj}-\phi ^{;ij})\right)\labell{first111}
\eeqa
where the semicolons are used to denote the covariant differentiation. 

The ${\alpha'}^2$ corrections to the D-brane action for one RR, one NSNS and one open string
NS states have been found from the low energy of the string amplitude \cite{Becker:2011ad, Komeil:2012np}. These corrections for the RR field strength $F^{(p)}$ are:
\beqa
\frac{\pi^2\alpha'^2T_p}{2!(p-2)!}\int d^{p+1}x\,\eps^{a_0\cdots a_p}
{ F}^{(p)}_{ija_3\cdots a_p}\left(\cR_{aa_0}{}^{ij,a}( 2\pi\alpha'f_{a_1a_2})-H_{a_0a_1a}{}^{,ai}(2\pi\alpha'\Phi^{j}{}_{,a_2})\right). \nonumber
\labell{first112}
\eeqa

where it has been extended $2\pi\alpha' f$ to gauge invariant $B+2\pi\alpha' f$ combination\cite{Garousi:2010bm}

The second term which involve the transverse scalar field, are exactly reproduced by the pull-back operator in \reef{first111} in the static gauge \cite{Komeil:2012np}. The first term should be added to the D-brane action at order $\alpha'^2$. Extending the first coupling above to be covariant under the coordinate transformations and invariant under the B-field and RR gauge transformations, one finds the following nonlinear couplings at order $\alpha'^2$:
\beqa
\!\!\!\!\!\!\!\! \frac{\pi^2\alpha'^2T_p}{2!(p-2)!} \int d^{p+1}x\,\eps^{a_0\cdots a_p}
{\tilde{F}}^{(p)}_{ija_3\cdots a_p}[2R_{aa_0}{}^{ij;a}(B_{a_1a_2}+2\pi\alpha'f_{a_1a_2}) ].
\labell{first113}
\eeqa
We calculate the field theory amplitudes by using the above couplings for the scattering of one RR $(p-1)$-form with  three and two transverse index and two NSNS states at order $\alpha'^2$. We then compare the results with low energy limit of string amplitudes at order $\alpha'^2$ and corresponding T-dual multiplets that has been found in \cite{Komeil:2012np}.  Using S-matrix method in details, the amplitude of one RR $(p-3)$-form with  one transverse index, and two arbitrary B-fields has been studied in \cite{Garousi:2010bm}. Here we perform similar computation for the amplitude of RR $(p-1)$-form in details and show that all structures of this amplitude, regardless the overall factor, can be produced by T-duality that has been found in \cite{Komeil:2012np}.

{It has been shown in \cite{Komeil:2012np} that the T-duality at linear order should appear in the amplitudes through the associated Ward identities. Scattering amplitude at any loop order should satisfy the T-dual Ward identity. This classifies the tree-level amplitudes into T-dual multiplets. Each multiplet includes the scattering amplitudes which interchange under the linear T-duality transformations.}



An outline of the paper is as follows: In section 2.1, we examine the calculation of string amplitude of one RR and two NSNS vertex operators in superstring theory. In section 2.2.1, we perform the calculation in full details for $C^{(p-1)}_{ijk}$ and expand the amplitude at low energy. Regarding the low energy expansion of integral term in amplitude, we show that there is neither  contact term nor massless open string pole at order $O(\alpha'^2)$ which is consistent with the couplings \reef{first111}. In section 2.2.2, we perform the same calculation  for $C^{(p-1)}_{ij}$. We find that the contact terms and the massless open string pole of the field theory, which have been found in \cite{Komeil:2012np}, are reproduced exactly by this amplitude at order $O(\alpha'^2)$. In fact, by using the standard S-matrix method, we find all elements of S-matrix in proposal case that found by T-duality in \cite{Komeil:2012np}.

\section{ String scattering amplitude } \label{intro2}
 
The S-matrix method is a very important tool in superstring theory to discover new couplings between strings. 
It has been shown that the corrections in field theory might have been obtained in $\alpha'$ by standard S-matrix method. A great deal of effort for the understanding of scattering amplitudes has been made \cite{Garousi:1998fg}.

The couplings in \reef{first111} can be confirmed by the scattering amplitude of one RR and two NSNS states that has been found by the T-duality in \cite{Komeil:2012np}. We reproduce these coupling by explicit calculations of S-matrix. 



Some of the couplings resulting from the consistency of the Chern-Simons action at order $O(\alpha'^2)$ with the linear T-duality transformations  confirm by the calculation of three closed string scattering amplitudes \cite{Becker:2010ij,Garousi:2010bm,Garousi:2011bm}. In this work, we are interested in finding all  couplings that string theory produces for  the RR potential $C^{(p-1)}$ which carries any world volume indices.


  To calculate a S-matrix element, one needs to choose the picture of the vertex operators appropriately. The sum of the superghost charge must be -2 for the disk level amplitude. The tree level scattering amplitude of one RR and two NSNS states on the world-volume of a D$_p$-brane has been studied in different vertex pictures  \cite{Becker:2010ij,Becker:2011ad,Garousi:2010bm,Garousi:2011bm}. We are interested to evaluate the amplitude of one RR, one graviton and one B-field in the picture that the symmetry between two NSNS states is manifest from the beginning. Hence, we work in $(-1/2,-3/2)$ picture  for RR state  \cite{Billo:1998vr} and the two NSNS vertex operators in $(0,0)$ picture. It has been shown that the final result is independent of the choice of the picture.

We work in the RNS worldsheet formalism, with the closed string vertex operators being constructed out of bosons $X^{\mu}(z), X^{\mu}(\bz)$ and fermions $\psi^{\mu}(z), \psi^{\mu}(\bz)$ as well as the picture ghosts $\phi(z)$ and $\phi(\bz)$.  We will now perform the computation for interested distribution of picture.

The string scattering amplitude is given by the following correlation function:
\beqa
\cA&\sim&<V_{RR}^{(-1/2,-3/2)}(\veps_1^{(n)},p_1)V_{NSNS}^{(0,0)}(\veps_2,p_2)V_{NSNS}^{(0,0)}(\veps_3,p_3)>\labell{amp2}
\eeqa
Using the doubling trick \cite{Garousi:1996ad}, explicitly we have:\footnote{Our conversions set $\alpha'=2$ in the string theory calculations.}
\beqa
V_{RR}^{(-1/2,-3/2)}&\!\!\!\!\!=\!\!\!\!\! &(P_-H_{1(n)}M_p)^{AB}\int d^2z_1:e^{-\phi(z_1)/2}S_A(z_1)e^{ip_1\cdot X}:e^{-3\phi(\bz_1)/2}S_B(\bz_1)e^{ip_1\cdot D\cdot  X}:\nonumber\\
V_{NSNS}^{(0,0)}&\!\!\!\!\!=\!\!\!\!\!&(\veps_2\inn D)_{\mu_3\mu_4}\int d^2z_2:(\prt X^{\mu_3}+ip_2\inn\psi\psi^{\mu_3})e^{ip_2\cdot X}:(\prt X^{\mu_4}+ip_2\inn D\inn\psi\psi^{\mu_4})e^{ip_2\cdot D\cdot X}:\nonumber\\
V_{NSNS}^{(0,0)}&\!\!\!\!\!=\!\!\!\!\!&(\veps_3\inn D)_{\mu_5\mu_6}\int d^2z_3:(\prt X^{\mu_5}+ip_3\inn\psi\psi^{\mu_5})e^{ip_3\cdot X}:(\prt X^{\mu_6}+ip_3\inn\nonumber D\inn\psi\psi^{\mu_6})e^{ip_3\cdot D\cdot X}:
\eeqa

where the NSNS vertex operators include polarizations $\veps_i$  and momenta $p_i$ ,$i=2,3$. The RR vertex operator has momentum  $p_1$ and antisymmetric polarization $\veps_1$. The indices $A,B,\cdots$ are the Dirac spinor indices and  $P_-=\frac{1}{2}(1-\gamma_{11})$ is the chiral projection operator, and
\beqa
H_{1(n)}&=&\frac{1}{n!}\veps_{1\mu_1\cdots\mu_{n}}\gamma^{\mu_1}\cdots\gamma^{\mu_{n}}\nonumber\\
M_p&=&\frac{\pm 1}{(p+1)!} \eps_{a_0 \cdots a_p} \ga^{a_0} \cdots \ga^{a_p}\labell{H1}
\eeqa
where $\eps$ is the volume $(p+1)$-form of the $D_p$-brane. Here the matrix $D_{\mu\nu}$ is a diagonal matrix that agrees with $\eta_{\mu\nu}$ in directions along the
brane (Neumann boundary conditions) and with $-\eta_{\mu\nu}$  in directions normal to the brane (Dirichlet boundary conditions). In this notation, $D^{\mu i}=-\delta^{\mu i},  D^{\mu a}=\delta ^{\mu a}$ and thus $D_{\mu\nu}=V_{\mu\nu}-N_{\mu\nu}$.

After performing the correlators, we will remove the volume of $SL(2,R)$ group which is the conformal symmetry of the  upper half $z$-plane \cite{Garousi:2010bm, Garousi:2011bm}. The amplitude \reef{amp2}  can be written  as 
\beqa
\cA&\sim&\frac{1}{2}(H_{1(n)}M_p)^{AB}(\veps_2\inn D)_{\mu_3\mu_4}(\veps_3\inn D)_{\mu_5\mu_6}\int d^2z_1d^2z_2d^2z_3\, (z_{1\bar 1})^{-3/4}\nonumber\\
&&(\sum_{i = 1 }^ {16}b_{i}) ^{\mu_3\mu_4\mu_5\mu_6}_{AB}\delta^{p+1}(p_1^a+p_2^a+p_3^a)+(2\leftrightarrow 3)\labell{A}
\eeqa

where we use the standard world-sheet propagators. $z_{i\bar j}=z_i-\bz_j$, and 
\begin{eqnarray}\labell{bi}
(b_1)^{\mu_3\mu_4\mu_5\mu_6}_{AB}&\!\!\!\!=\!\!\!\!&<:S_A(z_1):S_B(\bz_1):>g_1^{\mu_3\mu_4\mu_5\mu_6}\cr
(b_2)^{\mu_3\mu_4\mu_5\mu_6}_{AB}&\!\!\!\!=\!\!\!\!&(ip_2)_{\beta_1}<:S_A:S_B:\psi^{\beta_1}\psi^{\mu_3}:>g_2^{\mu_4\mu_5\mu_6}\cr
(b_3)^{\mu_3\mu_4\mu_5\mu_6}_{AB}&\!\!\!\!=\!\!\!\!&(ip_2\inn D)_{\beta_1}<:S_A:S_B:\psi^{\beta_1}\psi^{\mu_4}:>g_3^{\mu_3\mu_5\mu_6}\cr
(b_4)^{\mu_3\mu_4\mu_5\mu_6}_{AB}&=&(ip_3)_{\beta_1}<:S_A:S_B:\psi^{\beta_1}\psi^{\mu_5}:>g_4^{\mu_3\mu_4\mu_6}\cr
(b_5)^{\mu_3\mu_4\mu_5\mu_6}_{AB}&=&(ip_3\inn D)_{\beta_1}<:S_A:S_B:\psi^{\beta_1}\psi^{\mu_6}:>g_5^{\mu_3\mu_4\mu_5}\nonumber\\
(b_6)^{\mu_3\mu_4\mu_5\mu_6}_{AB}&\!\!\!\!=\!\!\!\!&(ip_2)_{\beta_1}(ip_2\inn D)_{\beta_2}<:S_A:S_B:\psi^{\beta_1}\psi^{\mu_3}:\psi^{\beta_2}\psi^{\mu_4}:>g_6^{\mu_5\mu_6}\\
(b_7)^{\mu_3\mu_4\mu_5\mu_6}_{AB}&\!\!\!\!=\!\!\!\!&(ip_2)_{\beta_1}(ip_3)_{\beta_2}<:S_A:S_B:\psi^{\beta_1}\psi^{\mu_3}:\psi^{\beta_2}\psi^{\mu_5}:>g_7^{\mu_4\mu_6}\nonumber\\
(b_8)^{\mu_3\mu_4\mu_5\mu_6}_{AB}&\!\!\!\!=\!\!\!\!&(ip_2)_{\beta_1}(ip_3\inn D)_{\beta_2}<:S_A:S_B:\psi^{\beta_1}\psi^{\mu_3}:\psi^{\beta_2}\psi^{\mu_6}:>g_8^{\mu_4\mu_5}\nonumber\\
(b_9)^{\mu_3\mu_4\mu_5\mu_6}_{AB}&=&(ip_2\inn D)_{\beta_1}(ip_3)_{\beta_2}<:S_A:S_B:\psi^{\beta_1}\psi^{\mu_4}:\psi^{\beta_2}\psi^{\mu_5}:>g_9^{\mu_3\mu_6}\nonumber\\
(b_{10})^{\mu_3\mu_4\mu_5\mu_6}_{AB}&\!\!\!\!=\!\!\!\!&(ip_2\inn D)_{\beta_1}(ip_3\inn D)_{\beta_2}<:S_A:S_B:\psi^{\beta_1}\psi^{\mu_4}:\psi^{\beta_2}\psi^{\mu_6}:>g_{10}^{\mu_3\mu_5}\nonumber\\
(b_{11})^{\mu_3\mu_4\mu_5\mu_6}_{AB}&=&(ip_3)_{\beta_1}(ip_3\inn D)_{\beta_2}<:S_A:S_B:\psi^{\beta_1}\psi^{\mu_5}:\psi^{\beta_2}\psi^{\mu_6}:>g_{11}^{\mu_3\mu_4}\nonumber\\
(b_{12})^{\mu_3\mu_4\mu_5\mu_6}_{AB}&\!\!\!\!=\!\!\!\!&(ip_2)_{\beta_1}(ip_2\inn D)_{\beta_2}(ip_3)_{\beta_3}<:S_A:S_B:\psi^{\beta_1}\psi^{\mu_3}:\psi^{\beta_2}\psi^{\mu_4}:\psi^{\beta_3}\psi^{\mu_5}:>g_{12}^{\mu_6}\nonumber\\
(b_{13})^{\mu_3\mu_4\mu_5\mu_6}_{AB}&\!\!\!\!=\!\!\!\!&(ip_2)_{\beta_1}(ip_2\inn D)_{\beta_2}(ip_3\inn D)_{\beta_3}<:S_A:S_B:\psi^{\beta_1}\psi^{\mu_3}:\psi^{\beta_2}\psi^{\mu_4}:\psi^{\beta_3}\psi^{\mu_6}:>g_{13}^{\mu_5}\nonumber\\
(b_{14})^{\mu_3\mu_4\mu_5\mu_6}_{AB}&=&(ip_2)_{\beta_1}(ip_3)_{\beta_2}(ip_3\inn D)_{\beta_3}<:S_A:S_B:\psi^{\beta_1}\psi^{\mu_3}:\psi^{\beta_2}\psi^{\mu_5}:\psi^{\beta_3}\psi^{\mu_6}:>g_{14}^{\mu_4}\nonumber\\
(b_{15})^{\mu_3\mu_4\mu_5\mu_6}_{AB}&=&(ip_2\inn D)_{\beta_1}(ip_3)_{\beta_2}(ip_3\inn D)_{\beta_3}<:S_A:S_B:\psi^{\beta_1}\psi^{\mu_4}:\psi^{\beta_2}\psi^{\mu_5}:\psi^{\beta_3}\psi^{\mu_6}:>g_{15}^{\mu_3}\nonumber\\
(b_{16})^{\mu_3\mu_4\mu_5\mu_6}_{AB}&\!\!\!\!=\!\!\!\!&(ip_2)_{\beta_1}(ip_2\inn D)_{\beta_2}(ip_3)_{\beta_3}(ip_3\inn D)_{\beta_4}\nonumber\\
&&\qquad\qquad\qquad\times<:S_A:S_B:\psi^{\beta_1}\psi^{\mu_3}:\psi^{\beta_2}\psi^{\mu_4}:\psi^{\beta_3}\psi^{\mu_5}:\psi^{\beta_4}\psi^{\mu_6}:>g_{16}\nonumber
\end{eqnarray}
where $g$'s are  the correlators  of $X$'s which can easily be performed using the standard world-sheet propagators, and  
the correlator of $\psi$ can be calculated using the  Wick-like rule \cite{Liu:2001qa}.

If the symmetry of two NSNS polarizations are similar then ten independent $b_i$ contribute to the amplitude as \cite{Garousi:2010bm,Garousi:2011bm}. Two NSNS polarizations in this work have different symmetry and sixteen independent $b_i$ contribute to the amplitude. By considering the relations \reef{H1} and \reef{bi} in amplitude \reef{A}, one can find that the scattering amplitude involves the following trace of gamma matrices :
\beqa
T(n,p,m) =\frac{1}{n!(p+1)!}\veps_{1\nu_1\cdots \nu_{n}}\eps_{a_0\cdots a_p}A_{[\alpha_1\cdots \alpha_m]}\Tr(\gamma^{\nu_1}\cdots \gamma^{\nu_{n}}\gamma^{a_0}\cdots\gamma^{a_p}\gamma^{\alpha_1\cdots \alpha_m})\labell{relation1}
 \eeqa
 
where $A_{[\alpha_1\cdots \alpha_m]}$ is an antisymmetric combination of the momenta and/or the polarizations of the NSNS states\cite{Garousi:2010bm, Garousi:2011bm}. 

 It could be verified that the trace \reef{relation1} is non-zero only  for $n=p-3,\, n=p-1,\, n=p+1,\, n=p+3,\, n=p+5$. The case $n=p-3$ is studied in \cite{Garousi:2010bm} and \cite{Garousi:2011bm} where the RR potential carries transverse indices and world volume indices, respectively.

 We are interested in the case $n=p-1$. In this case the only RR potentials that  lead to non zero amplitudes are that carried zero, one, two and three transvers indices. In this paper, we consider $(\veps_1^{(p-1)})^{ijk}$, $(\veps_1^{(p-1)})^{ij}$ and $(\veps_1^{(p-1)})^{i}$. The result can easily be extended to the RR  $n$-form by contracting its indices with the world volume form.   For the cases that we will evaluate, the trace relation \reef{relation1} gives non-zero result only for the following $T(n,p,m)$:
 \beqa
T(3,4,8)\quad\quad\quad T(2,3,6)\quad\quad\quad T(1,2,4)\nonumber
\eeqa 

So it could be concluded easily that $b_1,\, b_2$ and $b_3$ in \reef{bi} have no contribution to the amplitude. Let us begin with the RR potential $(\veps_1^{(p-1)})^{ijk}$. 

 \subsection{The amplitude of one $RR$ $(p-1)$-form  with three transvers indices and two $NSNS$.}
For this RR potential $n=3$, and from the relation $n=p-1$ one gets $p=4$. The trace  \reef{relation1} is non-zero only for $m=8$. It becomes 
\beqa
T(3,4,8)&=&32\frac{8!}{3!5!}\veps_1^{ijk}\eps^{a_0\cdots a_4}A_{[ijka_0\cdots a_4]}
\eeqa

where 32 is the trace of the $32\times 32$ identity matrix.
Since $m=8$,  only the $\psi$ correlator in $b_{16}$ have non-zero contribution to the amplitude \reef{amp2}. The $X$ correlator in  $b_{16}$ is
\beqa
g_{16}&\!\!\!\!=\!\!\!\!&|z_{12}|^{2p_1\cdot p_2}|z_{13}|^{2p_1\cdot p_3}|z_{23}|^{2p_2\cdot p_3}|z_{1\bar2}|^{2p_1\cdot D\cdot p_2}|z_{1\bar3}|^{2p_1\cdot D\cdot p_3}|z_{2\bar3}|^{2p_2\cdot D\cdot p_3} \nonumber\\
&&\times(z_{1\bar1})^{p_1\cdot D\cdot p_1}(z_{2\bar2})^{p_2\cdot D\cdot p_2}(z_{3\bar3})^{p_3\cdot D\cdot p_3}(i)^{p_1\cdot D\cdot p_1+p_2\cdot D\cdot p_2+p_3\cdot D\cdot p_3}
\eeqa

one can easily check that the integrand is  invariant under the $SL(2,R)$ transformation. So we can map the results to disk with unit radius. To fix this symmetry, we then set \cite{Craps:1998fn} $z_1=0$. The  correlator $b_{16}$ then becomes
 \beqa
|z_2| ^{2p_1\cdot p_2}|z_3|
^{2p_1\cdot p_3} \left(1-|z_2|^2 \right)^{p_2\cdot D\cdot p_2} \left(1-|z_3|^2\right)^{p_3\cdot D\cdot p_3}| z_2-z_3| ^{2p_2\cdot p_3}| 1-z_2\bar
z_3| ^{2p_2\cdot D\cdot p_3}\equiv K.\labell{K}
\eeqa
 
 Replacing in \reef{A} the above  $X$-correlator and the $\psi$-correlator  from Wick-like rule, one finds
 \beqa
{\cA}&\sim&2(\veps_1^{(p-1)})_{ijk}{}^{a_5\cdots a_p}\eps_{a_0\cdots a_p}p_2^{a_0}p_3^{a_1}p_2^ip_3^j\bigg((\veps_2^S)^{a_3k}(\veps_3^A)^{a_4a_2}+(2\leftrightarrow 3)\bigg)\nonumber\\
&&\times\cI_1\delta(p_1^a+p_2^a+p_3^a) \labell{2A3}
\eeqa
 
The amplitude for two graviton or two $B$-field is zero.This amplitude has been found as the second component of a T-dual multiplet (equation (15) in  \cite{Komeil:2012np}).

 The conservation of momentum is understood in all amplitudes in this paper. The integral in the above amplitude is
\beqa
\cI_{1}&=&\int_{|z_i|\leq 1} d^2z_2d^2z_3\frac{K}{|z_2|^2|z_3|^2}\labell{2I1}
\eeqa

To fix completely the $SL(2,R)$ symmetry, we then set \cite{Becker:2010ij} polar coordinates $z_i = r_i e^{ i \theta_i}$, $i=2,3$. Since the integrant depends only on $\theta_2-\theta_3$, one of the integrals can be
explicitly performed. To study the low energy limit of the amplitude \reef{2A3}, one has to expand $\cI_1$  at the low energy. We Taylor expand the integral in $r_2$ and $r_3$ using
\beqa\labell{TE}
 \frac{1}{(1-x)^m}=\sum_{n=0}^\infty\left(\begin{array}{c}
 m+n-1 \\n \\
 \end{array}
 \right)x^n
 \eeqa

The integration over $\theta_2-\theta_3$ produce the Kronecker delta function and the integrals over the radial coordinates are a set of infinite sums
\beqa 
\sum_{n_i = 0 }^ \infty && \left( \frac{1}{s+n_1+n_3+n_5} +\frac{1}{t+n_2+n_3+n_5} \right) \frac{\delta_{n_3-n_4+n_5-n_6,0}} {s+t+u+n_1+n_2+n_5 +n_6} \nonumber\\ \cr
&&\left(\begin{array}{c} -p-1+n_1 \\n_1 \\ \end{array}\right) \left(\begin{array}{c} -q-1+n_2 \\ n_2\\ \end{array}\right)\left(\begin{array}{c}-u-1+n_3 \\ n_3 \\ \end{array}\right)\cr
&& \left(\begin{array}{c} -u-1+n_4 \\ n_4 \\ \end{array}\right)\left(\begin{array}{c} -v-1+n_5 \\ n_5 \\ \end{array}\right) \left(\begin{array}{c} -v-1+n_6 \\ n_6\\ \end{array}\right)  \labell{appendix1}
\eeqa

where we have used the following definitions for the mandelstam variables:
\beqa
 s=p_1\cdot p_2 \quad &&; \quad t=p_1\cdot p_3 \quad ; \quad u=p_2\cdot p_3 \cr
 p=p_2\cdot D \cdot p_2 \quad &&\quad ; \quad q=p_3\cdot D \cdot p_3 \quad ; \quad v=p_2\cdot D \cdot p_3. 
\eeqa

Using \reef{appendix1} it is possible to show that asymptotically in the region of small
momenta the following expansions hold up to terms quadratic in momenta\footnote{In scattering theory, the number of momenta are related to the degree of $\alpha'$, so one can expand the integrals up to arbitrary degree of $\alpha'$.}
\beqa
I_{1}&=&\frac{\pi^2}{t}\bigg(\frac{1}{s+t+u}-\frac{\pi^2}{6}p\bigg)+\frac{\pi^2}{s}\bigg(\frac{1}{s+t+u}-\frac{\pi^2}{6}q\bigg).\labell{I1L}
\eeqa

The expansion of this integral is found in \cite{Garousi:2010bm} in a restricted kinematic setup. Since the amplitude considered in this paper has no massless pole in the $u$ and $v$-channel and also the kinematic factor $\veps_1^{ij}\eps^{a_0\cdots a_5}A_{[ija_0\cdots a_5]}$ does not have any term proportional to   $p_2\inn p_3$ and $p_2\inn D\inn p_3$ . So the integral in \reef{2A3} has no $u$ and $v$-channel poles. Hence,
 it is safe to restrict the Mandelstam variables in \reef{K} to $u=v=0$. By these consideration the expansion \reef{I1L} is exactly equal to the results in \cite{Garousi:2010bm}.
 
All terms in the above expansion are closed string poles. This  is consistent with the field theory calculation in the previous section that there is no massless open string channel for $(\veps_1^{(p-1)})^{ijk}$.

\subsection{The amplitude of one $RR$ $(p-1)$-form  with two transvers indices and two $NSNS$.}
In this case, $n=2$ and from the relation $n=p-1$, it could be found that $p=3$. Since the index of the RR potential is transverse and the indices of the volume form are the world-volume indices, one finds that the trace relation \reef{relation1} is non-zero only for $m=6$. The trace in this case becomes
\beqa
T(2,3,6)&=&32\frac{6!}{2!4!}\veps_1^{ij}\eps^{a_0\cdots a_3}A_{[ia_0\cdots a_3]}
\eeqa
 
  The $\psi$ correlators in $b_{12},\, b_{13},b_{14},\, b_{15}$ and $b_{16}$ have non-zero contributions. Using the on-shell condition $\veps_2\cdot p_2=\veps_3\cdot p_3=0$ and conservation of momentum along the world volume, one can find the $X$ correlators corresponding to these $b_{i}$'s , \ie,
\beqa
 g_{12}^{\mu_6}&=&  \frac{iK}{z_{3\bar3}}\left(\frac{p_1^{\mu _6}z_{31}}{z_{1\bar3}}+\frac{p_2^{\mu _6}z_{32}}{z_{2\bar3}}+\frac{(p_1\inn D)^{\mu _6}z_{3\bar1}}{{z}_{\bar1\bar3}}+\frac{(p_2\inn D)^{\mu _6}z_{3\bar2}}{{z}_{\bar2\bar3}}\right)\cr
g_{13}^{\mu_5}&=&  \frac{iK}{z_{\bar3 3}}\left(\frac{p_1^{\mu _5}z_{\bar3 1}}{z_{13}}+\frac{p_2^{\mu _5}z_{\bar3 2}}{z_{23}}+\frac{(p_1\inn D)^{\mu _5}z_{\bar3\bar1}}{{z}_{\bar1 3}}+\frac{(p_2\inn D)^{\mu _5}z_{\bar3\bar2}}{{z}_{\bar2 3}}\right)\cr 
g_{14}^{\mu_4}&=&  \frac{iK}{z_{2\bar2}}\left(\frac{p_1^{\mu _4}z_{21}}{z_{1\bar2}}+\frac{p_3^{\mu _4}z_{23}}{z_{3\bar2}}+\frac{(p_1\inn D)^{\mu _4}z_{2\bar1}}{{z}_{\bar1\bar2}}+\frac{(p_3\inn D)^{\mu _4}z_{2\bar3}}{{z}_{\bar3\bar2}}\right)\\\labell{2g}
g_{15}^{\mu_3}&=&  \frac{iK}{z_{\bar2 2}}\left(\frac{p_1^{\mu _3}z_{\bar2 1}}{z_{12}}+\frac{p_3^{\mu _3}z_{\bar2 3}}{z_{32}}+\frac{(p_1\inn D)^{\mu _3}z_{\bar2\bar1}}{{z}_{\bar1 2}}+\frac{(p_3\inn D)^{\mu _3}z_{\bar2\bar3}}{{z}_{\bar3 2}}\right)\cr 
g_{16}&\equiv& K\nonumber 
\eeqa

where $K$ is given in \reef{K}. 

We note that all terms behave similarly  under the $SL(2,R)$ transformation. Writing the sub-amplitudes  ${\cal A}_i$ in \reef{A} corresponding to $b_i$,  the sub-amplitudes corresponding to $b_{12},\, b_{13},b_{14},\, b_{15}$ are
\beqa
{\cal A}_{12}&\!\!\!\!\!\sim\!\!\!\!\!& 16\veps_1^{ij}\eps^{a_0\cdots a_3}(D\inn\veps_3^T)_{\mu_6a_1}(\veps_2)_{a_2j}(p_2)_{i}(p_2)_{a_3}(p_3)_{a_0}\cr
&&\int d^2z_1d^2z_2d^2z_3\frac{z_{1\bar1}K}{|z_{21}|^2|z_{2\bar1}|^2z_{31}z_{3\bar1}z_{3\bar3}}\left(\frac{p_1^{\mu _6}z_{31}}{z_{1\bar3}}+\frac{p_2^{\mu _6}z_{32}}{z_{2\bar3}}+\frac{(p_1\inn D)^{\mu _6}z_{3\bar1}}{{z}_{\bar1\bar3}}+\frac{(p_2\inn D)^{\mu _6}z_{3\bar2}}{{z}_{\bar2\bar3}}\right)\cr
{\cal A}_{13}&\!\!\!\!\!\sim\!\!\!\!\!& -16\veps_1^{ij}\eps^{a_0\cdots a_3}(\veps_3)_{\mu_5a_1}(\veps_2)_{a_2j}(p_2)_{i}(p_2)_{a_3}(p_3)_{a_0}\nonumber\\
&&\int d^2z_1d^2z_2d^2z_3\frac{z_{1\bar1}K}{|z_{21}|^2|z_{2\bar1}|^2z_{\bar31}z_{\bar3\bar1}z_{3\bar3}}\left(\frac{p_1^{\mu _5}z_{\bar3 1}}{z_{13}}+\frac{p_2^{\mu _5}z_{\bar3 2}}{z_{23}}+\frac{(p_1\inn D)^{\mu _5}z_{\bar3\bar1}}{{z}_{\bar1 3}}+\frac{(p_2\inn D)^{\mu _5}z_{\bar3\bar2}}{{z}_{\bar2 3}}\right)\nonumber\\
{\cal A}_{14}&\!\!\!\!\!\sim\!\!\!\!\!& 8\veps_1^{ij}\eps^{a_0\cdots a_3}(\veps_3)_{a_0a_2}(p_3)_{j}(p_3)_{a_1}\bigg[(D\inn\veps_2^T)_{\mu_4i}(p_2)_{a_3}-(D\inn\veps_2^T)_{\mu_4a_3}(p_2)_{i}\bigg]\nonumber\\
&&\int d^2z_1d^2z_2d^2z_3\frac{z_{1\bar1}K}{|z_{31}|^2|z_{3\bar1}|^2z_{21}z_{2\bar1}z_{2\bar2}}\left(\frac{p_1^{\mu _4}z_{21}}{z_{1\bar2}}+\frac{p_3^{\mu _4}z_{23}}{z_{3\bar2}}+\frac{(p_1\inn D)^{\mu _4}z_{2\bar1}}{{z}_{\bar1\bar2}}+\frac{(p_3\inn D)^{\mu _4}z_{2\bar3}}{{z}_{\bar3\bar2}}\right)\nonumber\\
{\cal A}_{15}&\!\!\!\!\!\sim\!\!\!\!\!& 8\veps_1^{ij}\eps^{a_0\cdots a_3}(\veps_3)_{a_0a_2}(p_3)_{a_1}(p_3)_{j}\bigg[(\veps_2)_{\mu_3i}(p_2)_{a_3}-(\veps_2)_{\mu_3a_3}(p_2)_{i}\bigg]\nonumber\\
&&\int d^2z_1d^2z_2d^2z_3\frac{z_{1\bar1}K}{|z_{31}|^2|z_{3\bar1}|^2z_{\bar31}z_{\bar2\bar1}z_{\bar2 2}}\left(\frac{p_1^{\mu _3}z_{\bar2 1}}{z_{12}}+\frac{p_3^{\mu _3}z_{\bar2 3}}{z_{32}}+\frac{(p_1\inn D)^{\mu _3}z_{\bar2\bar1}}{{z}_{\bar1 2}}+\frac{(p_3\inn D)^{\mu _3}z_{\bar2\bar3}}{{z}_{\bar3 2}}\right)\nonumber\\\labell{A1215}
\eeqa

It is obvious that the above amplitudes are non-zero for symmetric polarization tensor $\veps_2^S$ and antisymmetric polarization tensor $\veps_3^A$ . There are similar sub-amplitudes as above  in  the $(2\leftrightarrow 3)$ part of the amplitude \reef{A} where the polarization tensors are $\veps_2^A$ and $\veps_3^S$.

 The above amplitudes produce structures which contain the contraction of $p_1$ and $NSNS$ polarizations. In the world volume contraction of $p_1$, the conservation of momentum along the brane could be used to write $p_1\inn V\inn \veps$ in terms of $p_2\inn V\inn \veps$ and $p_3\inn V\inn \veps$.
 
By using this consideration and on-shell condition, it could be found that the contraction of momenta with corresponding $NSNS$ polarizations in the transvers direction are not independent structures, \ie
\beqa
p_2\inn N\inn \veps_2=- p_2\inn V\inn \veps_2 \quad\quad\quad\quad\quad\quad p_3\inn N\inn \veps_3=- p_3\inn V\inn \veps_3\labell{2pa1}
\eeqa
Therefor, the independent structures that contain the contraction of momentum and polarization are
\beqa
p_1\inn N\inn \veps_2 ,\  p_3\inn N\inn \veps_2 ,\  p_2\inn V\inn \veps_2 ,\  p_3\inn V\inn \veps_2 ,\  p_1\inn N\inn \veps_3 ,\  p_2\inn N\inn \veps_3 ,\  p_2\inn V\inn \veps_3 ,\  p_3\inn V\inn \veps_3\nonumber
\eeqa

using this fact that there is no momentum conservation in the transverse subspace.

 We will see that there are some other contributions as $pp$ and $\veps\veps$ from  $b_{16}$. 
Let us evaluate the only structures in the sub-amplitude  ${\cal A}_{16}$ in \reef{A} in details.

 We finally will collect all terms come from all subamplitudes and will write them in form of a final amplitude.
  
   The contractions which produce structure $p\veps_3$  give the following contribution to the amplitude ${\cal A}_{16}$:
\beqa
{\cal A}_{16}(p\veps_3^A)&\sim&-8 \veps_1^{ij}\eps^{a_0\cdots a_3}\int d^2z_1d^2z_2d^2z_3\frac{z_{1\bar1}^2K}{|z_{21}|^2|z_{31}|^2|z_{2\bar 1}|^2|z_{3\bar 1}|^2}\\
&& \bigg(\cP(z_2,z_3) A_{1[ija_0\cdots a_3]} +\cP(\bar{z}_2,z_3)A_{2[ija_0\cdots a_3]}
+\cP(z_3,\bar{z}_3)  A_{3[ija_0\cdots a_3]}\nonumber\\
&&+\cP(z_2,\bar{z}_3) A_{4[ija_0\cdots a_3]}
+\cP(z_3,{\bz}_3)A_{5[ija_0\cdots a_3]}  + \cP(\bar{z}_2,\bar{z}_3)A_{6[ija_0\cdots a_3]}\bigg)\nonumber
\eeqa
where $\cP(z_i,z_j)$ is given by the Wick-like contraction
\beqa
\cP(z_i,z_j)\eta^{\mu\nu}&=&\widehat{[\psi^{\mu}(z_i),\psi^{\nu}(z_j)]}=\eta^{\mu\nu}{\frac {(z_{i}-z_1)(z_{j}-\bz_1)+(z_{j}-z_1)(z_{i}-\bz_1)}{(z_{i}-z_{j})(z_1-\bz_1)}}\labell{ppp}
\eeqa 
Using the fact that above kinematic factors ($A_i$, $i=1,..,6$) contract with $\veps_1^{ij}\eps^{a_0\cdots a_3}$, one observes  that there are 15 different terms in each case, however, 11 of them are zero and the other four terms are equal. They are simplified as
\beqa
A_{1[ija_0\cdots a_3]}&=&\frac{4}{15}(p_2\inn\veps_3^A)_{a_3}(\veps_2^S)_{ia_1}(p_2)_{a_2}(p_3)_{j}(p_3)_{a_0}\nonumber\\
A_{2[ija_0\cdots a_3]}&=&-\frac{4}{15}(p_2\inn D\inn\veps_3^A)_{a_3}(\veps_2^S)_{ia_1}(p_2)_{a_2}(p_3)_{a_0}(p_3)_{j}\nonumber\\
A_{3[ija_0\cdots a_3]}&=&\frac{4}{15}(p_3\inn D\inn\veps_3^A)_{a_3}(\veps_2^S)_{ia_1}(p_2)_j(p_2)_{a_0}(p_3)_{a_2}\nonumber\\
A_{4[ija_0\cdots a_3]}&=&-\frac{4}{15}(p_2\inn D\inn(\veps_3^A)^T)_{a_3}(\veps_2^S)_{ia_1}(p_2)_{a_2}(p_3)_{a_0}(p_3)_{j}\nonumber\\
A_{5[ija_0\cdots a_3]}&=&-\frac{4}{15}(p_3\inn D\inn(\veps_3^A)^T)_{a_3}(\veps_2^S)_{ia_1}(p_2)_j(p_2)_{a_0}(p_3)_{a_2}\nonumber\\
A_{6[ija_0\cdots a_3]}&=&\frac{4}{15}(p_2\inn(\veps_3^A)^T)_{a_3}(\veps_2^S)_{ia_1}(p_2)_{a_2}(p_3)_{a_0}(p_3)_{j}\nonumber
\eeqa
These factors are zero for $\veps_2^A$ and $\veps_3^S$.

The contractions which produce structure $p\veps_2^S$  give the following contribution to the amplitude ${\cal A}_{16}$:
\beqa
{\cal A}_{16}(p\veps_2^S)&\sim&-4 \veps_1^{ij}\eps^{a_0\cdots a_3}\int d^2z_1d^2z_2d^2z_3\frac{z_{1\bar1}^2K}{|z_{21}|^2|z_{31}|^2|z_{2\bar 1}|^2|z_{3\bar 1}|^2}\\
&& \bigg(\cP(z_2,z_3) A'_{1[ija_0\cdots a_3]} +\cP(\bar{z}_2,z_3)A'_{2[ija_0\cdots a_3]}
+\cP(z_3,\bar{z}_3)  A'_{3[ija_0\cdots a_3]}\nonumber\\
&&+\cP(z_2,\bar{z}_3) A'_{4[ija_0\cdots a_3]}
+\cP(z_3,{\bz}_3)A'_{5[ija_0\cdots a_3]}  +\cP(\bar{z}_2,\bar{z}_3)A'_{6[ija_0\cdots a_3]}\bigg)\nonumber
\eeqa
where 
\beqa
A'_{1[ija_0\cdots a_3]}&=&\frac{2}{15}[(p_3\inn\veps_2^S)_{i}(p_3)_{a_2}-(p_3\inn\veps_2^S)_{a_2}(p_3)_{i}](\veps_3^A)_{a_0a_1}(p_2)_{a_3}(p_2)_{j}\cr
A'_{2[ija_0\cdots a_3]}&=&\frac{2}{15}[(p_3\inn D\inn(\veps_2^S)^T)_{i}(p_3)_{a_2}+(p_3\inn D\inn(\veps_2^S)^T)_{a_2}(p_3)_{i}](\veps_3^A)_{a_0a_1}(p_2)_{a_3}(p_2)_{j}\cr
A'_{3[ija_0\cdots a_3]}&=&-\frac{2}{15}[(p_2\inn D\inn\veps_2^S)_{i}(p_2)_{a_2}+(p_2\inn D\inn\veps_2^S)_{a_2}(p_2)_{i}](\veps_3^A)_{a_0a_1}(p_3)_{a_3}(p_3)_{j}\cr
A'_{4[ija_0\cdots a_3]}&=&-\frac{2}{15}[(p_3\inn D\inn\veps_2^S)_{i}(p_3)_{a_2}+(p_3\inn D\inn\veps_2^S)_{a_2}(p_3)_{i}](\veps_3^A)_{a_0a_1}(p_2)_{a_3}(p_2)_{j}\nonumber\\
A'_{5[ija_0\cdots a_3]}&=&-\frac{2}{15}[(p_2\inn D\inn(\veps_2^S)^T)_{i}(p_2)_{a_2}+(p_2\inn D\inn(\veps_2^S)^T)_{a_2}(p_2)_{i}](\veps_3^A)_{a_0a_1}(p_3)_{a_3}(p_3)_{j}\nonumber\\
A'_{6[ija_0\cdots a_3]}&=&-\frac{2}{15}[(p_3\inn(\veps_2^S)^T)_{i}(p_3)_{a_2}-(p_3\inn(\veps_2^S)^T)_{a_2}(p_3)_{i}](\veps_3^A)_{a_0a_1}(p_2)_{a_3}(p_2)_{j}\nonumber
\eeqa
It is easy to verify that the above amplitude is zero, when NSNS polarization tensors are $\veps_2^A$ and $\veps_3^S$.

The $pp$ structures in ${\cal A}_{16}$ are found in terms of $p_2\inn D\inn p_3,\,  p_2\inn D\inn p_2,\,  p_3\inn D\inn p_3$  and $p_2\inn p_3$.
\beqa
{\cal A}_{16}(pp)&\sim&\frac{6!}{4!}\veps_1^{ij}\eps^{a_0\cdots a_3}\int d^2z_1d^2z_2d^2z_3\frac{z_{1\bar1}^2K}{|z_{21}|^2|z_{31}|^2|z_{2\bar 1}|^2|z_{3\bar 1}|^2}\\
&& \bigg(\cP(z_2,z_3) A_{7[ija_0\cdots a_3]} +\cP(z_2,\bar{z}_2)A_{8[ija_0\cdots a_3]}
+\cP(\bz_2,{z}_3)  A_{9[ija_0\cdots a_3]}\nonumber\\
&&+\cP(z_2,\bar{z}_3) A_{10[ija_0\cdots a_3]}
+\cP(z_3,{\bz}_3)A_{11[ija_0\cdots a_3]} +\cP(\bar{z}_2,\bar{z}_3)A_{12[ija_0\cdots a_3]}  \bigg) \nonumber
\eeqa
When both NSNS polarization tensors are symmetric or antisymmetric, some non-zero terms could be found. However, the integrand for those terms are pure imaginary which are  zero after integration. When the above kinematic factors ($A_i$, $i=7,..,12$) contain $\veps_2^S$ and $\veps_3^A$, they are
\beqa
A_{7[ija_0\cdots a_3]}&=&-A_{12[ija_0\cdots a_3]}=\frac{2}{15}(p_2\inn p_3)\big[(p_2)_{a_2}(p_3)_{j}-(p_2)_{j}(p_3)_{a_2}\big](\veps_2^S)_{ia_1}(\veps_3^A)_{a_0a_3}\nonumber\\
A_{8[ija_0\cdots a_3]}&=&-\frac{4}{15}(p_2\inn D\inn p_2)(p_3)_{a_2}(p_3)_{j}(\veps_2^S)_{ia_1}(\veps_3^A)_{a_0a_3}\nonumber\\
A_{9[ija_0\cdots a_3]}&=&-A_{10[ija_0\cdots a_3]}=-\frac{2}{15}(p_2\inn D\inn p_3)\big[(p_2)_{a_2}(p_3)_{j}+(p_2)_{j}(p_3)_{a_2}\big](\veps_2^S)_{ia_1}(\veps_3^A)_{a_0a_3}\nonumber\\
A_{11[ija_0\cdots a_3]}&=&-\frac{4}{15}(p_3\inn D\inn p_3)(p_2)_{a_2}(p_2)_{j}(\veps_2^S)_{ia_1}(\veps_3^A)_{a_0a_3}\nonumber
\eeqa

The another structure that contribute to the amplitude  ${\cal A}_{16}$ is $\veps_2\veps_3$:
\beqa
{\cal A}_{16}(\veps_2^S\veps_3^A)&\sim&8 \veps_1^{ij}\eps^{a_0\cdots a_3}\int d^2z_1d^2z_2d^2z_3\frac{z_{1\bar1}^2K}{|z_{21}|^2|z_{31}|^2|z_{2\bar 1}|^2|z_{3\bar 1}|^2}\\
&& \bigg(\cP(z_2,z_3) A'_{7[ija_0\cdots a_3]} +\cP(\bar{z}_2,z_3)A'_{8[ija_0\cdots a_3]}
+\cP(z_3,\bar{z}_3)  A'_{9[ija_0\cdots a_3]}\nonumber\\
&&+\cP(z_2,\bar{z}_3) A'_{10[ija_0\cdots a_3]}
  +\cP(\bar{z}_2,\bar{z}_3)A'_{11[ija_0\cdots a_3]}\bigg)\nonumber
\eeqa
where
\beqa
A'_{7[ija_0\cdots a_3]}&=&-A'_{11[ija_0\cdots a_3]}=\frac{4}{15}(p_2)_{a_2}(p_3)_{a_0}(p_2)_{i}(p_3)_{j}(\veps_2^S \inn \veps_3^A)_{a_1a_3}\nonumber\\
A'_{8[ija_0\cdots a_3]}&=&-A'_{10[ija_0\cdots a_3]}=-\frac{4}{15}(p_2)_{a_2}(p_3)_{a_0}(p_2)_{i}(p_3)_{j}(\veps_2^S \inn D \inn \veps_3^A)_{a_1a_3}\nonumber\\
A'_{9[ija_0\cdots a_3]}&=&0\nonumber
\eeqa

There are the contribution to the amplitude  ${\cal A}_{16}$ in which contain structures with any contraction. In fact, all momenta and both NSNS polarizations contract with RR polarization and volume $(p+1)$-form. In this structure the symmetric polarization tensor appears in the form of a trace.
\beqa
{\cal A}_{16}(no-contraction)&\sim&8 \veps_1^{ij}\eps^{a_0\cdots a_3}\int d^2z_1d^2z_2d^2z_3\frac{z_{1\bar1}^2K}{|z_{21}|^2|z_{31}|^2|z_{2\bar 1}|^2|z_{3\bar 1}|^2}\cP(z_2,{\bz}_2)\nonumber\\
&&\times \bigg(\frac{4}{15}\Tr(\veps_2^S \inn D)(p_2)_{a_3}(p_3)_{a_0}(p_2)_{i}(p_3)_{j}( \veps_3^A)_{a_1a_2}\bigg)
\eeqa
These two latter amplitudes are also zero for antisymmetric polarization tensor $\veps_2$ and symmetric polarization tensor $\veps_3$.

We perform this calculation to find all structures that contribute to the nonzero sub-amplitudes ${\cal A}_{12}, \ {\cal A}_{13},\ {\cal A}_{14}$ and ${\cal A}_{15}$. One can observe when both NSNS polarization tensors are symmetric or antisymmetric, the result is zero. The result for one graviton and one B-field is
\beqa
{\cal A}&\!\!\!\!\!\sim\!\!\!\!\!&  (\veps_1^{(p-1)})_{ij}{}^{a_4\cdots a_p}\eps_{a_0\cdots a_p}p_2^{a_3}\bigg[
2p_2^i p_3^{a_0}\bigg(2(\veps_2^S)^{a_2j}(p_1\inn N\inn\veps_3^A)^{a_1}+(\veps_2^A)^{a_1a_2}(p_1\inn N\inn\veps_3^S)^j\bigg)\cI_1\nonumber\\
&&-p_2^i p_3^{a_0}\bigg(2(\veps_2^S)^{a_2j}(p_2\inn N\inn \veps_3^A)^{a_1}+(\veps_2^A)^{a_1a_2}(p_2\inn N\inn\veps_3^S)^j\bigg)\cI_2\nonumber\\
&&-p_3^i p_3^{a_0}\bigg(2(\veps_2^S)^{a_2j}(p_2\inn V\inn\veps_3^A)^{a_1}+(\veps_2^A)^{a_1a_2}(p_2\inn V\inn\veps_3^S)^j\bigg)\cI_2\nonumber\\
&&+p_3^i p_2\inn V\inn p_3\bigg((\veps_2^S)^{a_2j}(\veps_3^A)^{a_0a_1}+(2\leftrightarrow 3)\bigg)\cI_2\labell{n=2polarization}\\
&&+p_2^i p_3^{a_0}\bigg(2(\veps_2^S)^{a_2j}(p_2\inn V\inn \veps_3^A)^{a_1}+(\veps_2^A)^{a_1a_2}(p_2\inn V\inn\veps_3^S)^j\bigg)\cI_3\nonumber\\
&&+p_3^i p_3^{a_0}\bigg(2(\veps_2^S)^{a_2j}(p_2\inn N\inn\veps_3^A)^{a_1}+(\veps_2^A)^{a_1a_2}(p_2\inn N\inn\veps_3^S)^j\bigg)\cI_3\nonumber\\
&&-p_3^i p_2\inn N\inn p_3\bigg((\veps_2^S)^{a_2j}(\veps_3^A)^{a_0a_1}+(2\leftrightarrow 3)\bigg)\cI_3\nonumber\\
&&-4p_2^i p_3^{a_0}\bigg(2(\veps_2^S)^{a_2j}(p_3\inn V\inn \veps_3^A)^{a_1}+(\veps_2^A)^{a_1a_2}(p_3\inn V\inn\veps_3^S)^j\bigg)\cI_4\nonumber\\
&&+2p_2^i p_3\inn V\inn p_3\bigg((\veps_2^S)^{a_2j}(\veps_3^A)^{a_0a_1}+(2\leftrightarrow 3)\bigg)\cI_4\nonumber\\
&&-2 p_2^ip_3^j p_3^{a_0} \bigg((\veps_2^S\inn V\inn\veps_3^A)^{a_2a_1}\cI_2-(\veps_2^S\inn N\inn\veps_3^A)^{a_2a_1}\cI_3\bigg)\nonumber\\
&&+ p_2^ip_3^j(\veps_3^A)^{a_0a_1}\bigg(\, (p_3\inn V\inn \veps_2^S)^{a_2}  \cI_2-\, (p_3\inn N\inn \veps_2^S)^{a_2}  \cI_3\bigg)\nonumber\\
&& -p_2^ip_3^j (\veps_2^A)^{a_1a_2}\bigg(2(p_1\inn N\inn\veps_3^S)^{a_0}\cI_1 -(p_2\inn N\inn\veps_3^S)^{a_0}\cI_2+(p_2\inn V\inn\veps_3^S)^{a_0}\cI_3\nonumber\\
&&-\Tr(\veps_3^S\inn D)p_3^{a_0}\cI_4\bigg)
\bigg]+(2\leftrightarrow 3)\nonumber
\eeqa
where  the function $\cI_1$ is the one appears in \reef{2I1} and $\cI_2, \cI_4$ are
\beqa
\cI_2&=&\int_{|z_i|\leq 1} d^2z_2d^2z_3\frac{1}{|z_2|^2|z_3|^2} \bigg(\frac{1-|z_2|^2|z_3|^2}{|1-z_2\bar z_3|^2}-\frac{|z_2|^2-|z_3|^2}{|z_2-z_3|^2}\bigg) \cal K\nonumber\\
\cI_4&=&-\int_{|z_i|\leq 1} d^2z_2d^2z_3\frac{1+|z_3|^2}{|z_2|^2|z_3|^2(1-|z_3|^2)} \cal K\labell{2I2I4}
\eeqa

The function $\cI_3$ is the same as the function $\cI_2$ in which the momentum labels 2 and 3 are exchanged. These functions satisfy the following relations:
\beqa
-2p_1\inn N\inn p_3\cI_1+2p_3\inn V\inn p_3\cI_4+p_2\inn N\inn p_3\cI_2-p_2\inn V\inn p_3\cI_3&=&0\nonumber\\
-2p_1\inn N\inn p_2\cI_1+2p_2\inn V\inn p_2\cI_7+p_2\inn N\inn p_3\cI_3-p_2\inn V\inn p_3\cI_2&=&0\labell{identity}
\eeqa
where the function $\cI_7$ is the same as the function $\cI_4$, in which the momentum labels 2 and 3 are exchanged.

 The amplitude  \reef{n=2polarization} satisfies the Ward identities associated with the symmetric and antisymmetric NSNS gauge transformations. It would satisfy the Ward identity associated with the RR gauge transformations when it combines with the amplitude \reef{2A3}, the amplitude of the RR $(p-1)$-form with one transverse index (see appendix) and RR $(p-1)$-form with any transverse index in which we are not interested in this work. 
 {This amplitude has been found in \cite{Komeil:2012np} as the sum of the second and the first components of the T-dual multiplets $\cA_2$ and $\cA'_2$ respectively (equation (34) in \cite{Komeil:2012np})}. The amplitude has been written in terms of H,\ie
\beqa
{\cA}&\sim &2(\veps_1^{(p-1)})_{ij}{}^{a_4\cdots a_p}\eps_{a_0\cdots a_p}\bigg[-H_3^{a_0a_1a_2}p_3^{j}\bigg(2p_2^{[i}(p_1\inn N\inn \veps_2^S)^{a_3]} \cI_1+p_2^{[i}(p_3\inn U\inn \veps_2^S)^{a_3]}  \nonumber\\
&& +[4p_2^{a_3}(p_2\inn V\inn\veps_2^S)^i-p_2^{a_3}p_2^i \Tr(\veps_2^S\inn D)-2p_2\inn V\inn p_2(\veps_2^S)^{a_3i}]\cI_7\bigg)\nonumber\\
&&-3p_2^ip_2^{a_3}(\veps_2^S)^{a_2j}\bigg(2(p_1\inn N\inn H_3)^{a_0a_1}\cI_1 +(p_2\inn W\inn H_3)^{a_0a_1}  -4(p_3\inn V\inn H_3)^{a_0a_1}\cI_4\bigg)\nonumber\\
&&+3p_3^{j}p_2^{a_3}\bigg(-(\veps_2^S)^{a_2i} (p_2\inn U\inn H_3)^{a_0a_1} +p_2^i(\veps_2^S \inn U\inn H_3)^{a_0a_1a_2}  \bigg)\nonumber\\
&&+ H_3^{a_0a_1a_2}p_2^j\bigg((\veps_2^S)^{ia_3}p_2\inn W\inn p_3\ -p_2^{a_3}(p_3\inn W\inn\veps_2^S)^i \bigg) \bigg]+(2\leftrightarrow 3)\labell{totalff}
\eeqa
where $H_i$ is the field strength of the $B_i$ polarization tensor, \eg
\beqa
H_i^{\mu\nu\alpha}&=&i[(\veps_i^A)^{\mu\nu}p_i^{\alpha}+(\veps_i^A)^{\alpha\mu}p_i^{\nu}+(\veps_i^A)^{\nu\alpha}p_i^{\mu}]
\eeqa
$U$ and $W$ have been defined in\cite{Komeil:2012np}  as $U\equiv V\cI_2-N\cI_3, W\equiv V\cI_3-N\cI_2$. Note that we have used the identity \reef{identity} to write the above amplitude in terms of field strength $H$. The amplitude does not satisfy the Ward identity corresponding to the graviton unless one rewrite the field strength $H$ in terms of $\veps^A$. Therefor, one can not write the amplitude in terms of  field strengths  $H$ and the curvature $R$. However, some of them can be written as $RH$. Because  the metric  appears in the effective action in the curvature tensor as well as in contracting the indices and in the definition of the covariant derivatives, one can not expect that all terms to be rewritten as $RH$.

To study the low energy limit of the amplitude, one has to expand $\cI_2$ and $\cI_4$ at the low energy as well as $\cI_1$ in \reef{I1L}. (Mapping the integrand in each of the integrals $\cI_2$ and $\cI_4$ to unit disk and fixing the $SL(2,R)$ symmetry as in section 2.2.1).  Taylor expansion of integrals $\cI_4$  in $r_2$ and $r_3$ using \reef{TE} and then the integrals over the radial coordinates contain following infinite sums
\beqa 
\sum_{n_i = 0 }^ \infty && \bigg[\left( \frac{1}{s+n_1+n_3+n_5} +\frac{1}{t+n_2+n_3+n_5} \right) \frac{1} {s+t+u+n_1+n_2+n_5 +n_6}+ \nonumber\\ \cr
&&\left( \frac{1}{s+1+n_1+n_3+n_5} +\frac{1}{t+n_2+n_3+n_5} \right) \frac{1} {s+t+u+1+n_1+n_2+n_5 +n_6}\bigg]\nonumber\\ 
&&\left(\begin{array}{c} -p+n_1 \\n_1 \\ \end{array}\right) \left(\begin{array}{c} -q-1+n_2 \\ n_2\\ \end{array}\right)\left(\begin{array}{c}-u-1+n_3 \\ n_3 \\ \end{array}\right)\cr
&& \left(\begin{array}{c} -u-1+n_4 \\ n_4 \\ \end{array}\right)\left(\begin{array}{c} -v-1+n_5 \\ n_5 \\ \end{array}\right) \left(\begin{array}{c} -v-1+n_6 \\ n_6\\ \end{array}\right)\delta_{n_3-n_4+n_5-n_6,0}  
\eeqa
One can find such infinite sums for the integral $\cI_2$.  Using these sums, in the region of small momenta the following expansions  can be found for these integrals:
\beqa
\cI_2&=&\frac{2 \pi ^2}{s (s+t)}-\frac{\pi ^4 q}{3 s}+...\cr && \cr
\cI_4&=&\frac{\pi ^4 p}{3 q}+\frac{\pi ^4 p}{6 t}+\frac{\pi ^4 t}{3s}+\frac{\pi ^4 q}{6 s}-\frac{2 \pi ^2}{q s}-\frac{\pi ^2}{s (s+t)}-\frac{\pi ^2}{t (s+t)}+....
\eeqa
the expansions hold up to terms quadratic in momenta. It is obvious from the above expansions and the expansion \reef{I1L} that $\cI_1$, $\cI_2$ and $\cI_3$ have only closed string poles while the integrals $\cI_4$ and $\cI_7$ have both open and closed poles. To find the open pole amplitude one has to consider the first two terms in the second line and the last term in the third line of \reef{totalff}. Hence, the amplitude \reef{totalff} has the following massless open string pole at order $O(\alpha'^2)$: 
\beqa
{\cal A}_{L-E}^{pole}&\sim&-4\pi^4\frac{p_3\inn V\inn p_3}{p_2\inn V\inn p_2}\eps_{a_0\cdots a_3} p_2^{a_2}p_3^{a_3}\bigg[-2(p_2\inn V\inn \veps_2^A)^{a_1}(p_3\inn N\inn \veps_1\inn N\inn \veps_3^S)^{a_0}\nonumber\\
&+&\bigg(p_2\inn N\inn \veps_1\inn N\inn p_3\, Tr(\veps_2^S\inn V)+2\,p_2\inn V\inn\veps_2^S\inn N\inn \veps_1 \inn N\inn p_3\bigg)(\veps_3^A)^{a_0a_1}\bigg]+(2\leftrightarrow 3).\labell{open2ij}
\eeqa
We will compare this amplitude with the corresponding feild theory amplitude in the next section.
The part of the third term in the third line and the third term in the second line of \reef{totalff} produce following contact term:
\beqa
4\pi^4 \eps_{a_0\cdots a_3} p_3^{a_3} p_3\inn V\inn p_3 \bigg[(p_3\inn N\inn \veps_1\inn N\inn \veps_3)^{a_0}(\veps_2^A)^{a_1a_2}+(p_3\inn N\inn \veps_1\inn N\inn \veps_2)^{a_0}(\veps_3^A)^{a_1a_2}\bigg]+(2\leftrightarrow 3).\nonumber
\eeqa
These contact terms contribute to the couplings \reef{first111} and \reef{first113}.  

\section{Field theory amplitude}
The scattering amplitude of one RR, one graviton and one B-field in the field theory is given by the following two Feynman amplitudes:
\beqa
{\cal A}^{A-pole}&=&V_a(\veps_2^A,A)G_{ab}(A) V_b(A,\veps_3^S,\veps_1)+(2\leftrightarrow 3)
\eeqa
where $A$ is the gauge field on the D$_p$-brane and
 \beqa
{\cal A}^{\Phi-pole}&=&V_m(\veps_2^S,\Phi)G_{mn}(\Phi) V_n(\Phi,\veps_3^A,\veps_1)+(2\leftrightarrow 3)
\eeqa
where $\Phi$ is the scalar field on the D$_p$-brane. In these amplitudes and in all amplitudes in this paper, the polarization of the RR field is given by $\veps_1$ and the polarizations of the graviton and B-field are given by $\veps^S,\, \veps^A$ respectively.

We assume that the RR potentials carry no world volume indices. Then the only non-zero vertex $ V_b(A,\veps^S,\veps_1)$ is given by the second term of \reef{first113} for $p=3$ and the vertex $ V_n(\Phi,\veps^A,\veps_1)$ arises from the second term in \reef{first112}:
 \beqa
V_b(A,\veps_3^S,\veps_1)&=&\frac{4(\pi\alpha')^3T_3}{3!}\eps_{a_0 a_1 a_{2}b}\veps_1^{ij}(\veps_3^S)^{a_0i}(p_3\inn V\inn p_3)p_3^{a_1}p_3^jp_2^{a_2}\labell{ahc}\nonumber\\
V_n(\Phi,\veps_3^A,\veps_1)&=&-\frac{(\pi\alpha')^2T_3}{2}\eps_{a_0\cdots a_3} (\veps_1)_{ij}p_2^{a_2}p_3^{a_3}p_3^i(p_3\inn V\inn p_3)(\veps_3^A)^{a_0a_1}\nonumber
\eeqa
Spacetime vectors project  into transverse and parallel subspace to the D$_p$-brane by the matrices  $N_{\mu\nu}$ and $V_{\mu\nu}$, respectively. $T_p$ is the tension of $D_p$-brane in the string frame
\beqa
T_p=\frac{1}{g_s(2\pi)^p(\alpha')^{(p+1)/2}}   . \labell{2T}\nonumber
\eeqa
  The gauge field propagator and the vertex $V_a(\veps^A,A)$ can be read from the DBI action \reef{DBI}, \ie
\beqa
G_{ab}(A)&=&\left(\frac{-i}{T_3(2\pi\alpha')^2}\right)\frac{\eta_{ab}}{p_2\inn V\inn p_2},\nonumber\\
V_a(\veps_2^A,A)&=&(2\pi\alpha')T_3(p_2\inn V\inn\veps_2^A)_a . \nonumber
\eeqa
The scalar field propagator and the vertex $V_a(\Phi,\veps^S)$ arises from the pull-back and the Taylor expansion
of the linear graviton in the DBI action, \ie
\beqa
G_{mn}(\Phi)&=&\frac{-i}{T_3}\frac{\eta_{mn}}{p_2\inn V\inn p_2} , \nonumber\\
V_m(\veps_2^S,\Phi)&=&-2T_3\bigg(p_2^iTr(\veps_2^S\inn V)+(p_2\inn V\inn\veps_2^S)^i\bigg) . \nonumber
\eeqa
The pole amplitude then becomes
\beqa
{\cal A}^{pole}&=&{\cal A}^{A-pole}+{\cal A}^{\Phi-pole}\nonumber\\
&=&-i\left(\frac{(\pi\alpha')^2T_3}{3}\right)\frac{p_3\inn V\inn p_3}{p_2\inn V\inn p_2}\eps_{a_0\cdots a_3} (\veps_1)_{ij}p_2^{a_2}p_3^{a_3}p_3^j\\\labell{open2ij}
&\times&\bigg[(p_2\inn V\inn \veps_2^A)^{a_1}(\veps_3^S)^{a_0i}+\bigg(3 p_2^iTr(\veps_2^S\inn V)+3(p_2\inn V\inn\veps_2^S)^i\bigg)(\veps_3^A)^{a_0a_1}\bigg]+(2\leftrightarrow 3).\nonumber
\eeqa
This amplitude has  six momenta in the numerator and two momenta in the denominator ($O(\alpha'^2)$).

The first term in \reef{first113} and the first term in \reef{first111} create contact terms at the order of  $O(\alpha'^2)$ that are simplified to
\beqa
{\cal A}^{contact}=-i\left(\frac{(\pi\alpha')^2T_3}{2!3!}\right)\eps_{a_0\cdots a_3} (\veps_1)_{ij}(\veps_3^A)^{a_2a_3}R_2^{a_1a_0ij}p_2\inn V\inn p_2+(2\leftrightarrow 3)\nonumber. 
\eeqa
The amplitude corresponding to the sum of ${\cal A}^{pole}$ and ${\cal A}^{contact}$  is the following:

\beqa
{\cal A}&=&-i\left(\frac{(\pi\alpha')^2T_3}{6}\right)\frac{p_3\inn V\inn p_3}{p_2\inn V\inn p_2}\eps_{a_0\cdots a_3}(\veps_1)_{ij} p_3^{j}\\
&&\times \bigg[p_3^{a_1}(\veps_3^S)^{a_0i}(p_2\inn V\inn H_2)^{a_2a_3}+2p_2^{a_2}\bigg(3 p_2^iTr(\veps_2^S\inn V)+(p_2\inn V\inn\veps_2^S)^i\bigg) H_3^{a_0a_1a_2}\bigg]+(2\leftrightarrow 3)\nonumber
\eeqa


The above feild theory open amplitude  is exactly the string low energy result \reef{open2ij} provided that the normalization factor of the string amplitude  \reef{totalff} is fixed at $ (3i\alpha'^2 T_p)/(4\pi^2)$. 

Therefor, by using the explicit calculation of string theory disk amplitude and compare with corresponding field theory evaluation, we examine the S-matrix of one RR $(p-1)$-form (with three and two transvers indices) and two NSNS states and also their corresponding couplings.  We illustrate the consistency between T-duality and explicit calculation of string scattering. At first, we show that the S-matrix that produced by T-dual ward identity is reproduced by three point amplitude and then we show this consistency between corresponding couplings.
We have performed the same steps for the case that the RR $(p-1)$-form carry one transvers index. We present this result in the appendix.


{\bf Acknowledgments}:  We would like to thank M.R.Garousi and A.Jalali for very valuable discussions and H.Zamani for very helpful conversations. This work is supported by University of Guilan.

\newpage
{\bf\Large {A\quad The amplitude of one $RR$ $(p-1)$-form  with one transvers index and two $NSNS$}}

In this appendix we calculate three-point amplitude \reef{A} for one RR potential $(C^{p-1})^{i}$ and two NSNS states. This scattering amplitude is zero for two graviton or two B-field vertex operators. Using the same steps as in sections 2.2.1 and 2.2.2, we find this result by explicitly calculating in $(-3/2,-1/2)$-picture. For this case, we have $n=1$ and $p=2$. The trace \reef{relation1} is non-zero only for T(1,2,4). The $\psi$ correlators in $b_{6},b_{7},b_{8},b_{9},b_{10},b_{11},b_{12},b_{13},b_{14},b_{15}$ and $b_{16}$ have non-zero contribution to the amplitude \reef{amp2}. 

Using the explicit calculation, one can find that the polarization of $RR$ $(p-1)$-form with one transvers index appears in amplitude in the form of following RR structures:
\beqa
&&\quad\quad(p\inn N\inn \veps_1)^{\mu}\quad , \quad(\veps^S \inn N\inn \veps_1)^{\mu\nu}\nonumber\\
&&(p\inn V\inn \veps^S \inn N\inn \veps_1)^{\mu}\quad , \quad(p\inn N\inn \veps^S \inn N\inn \veps_1)^{\mu}\nonumber\\
&&(p\inn N\inn \veps^A \inn N\inn \veps_1)^{\mu}\quad , \quad(p\inn V\inn \veps^A \inn N\inn \veps_1)^{\mu}\nonumber\\
&&(\veps_1\inn N\inn \veps^S \inn V\inn \veps^A)^{\mu\nu}\quad , \quad(\veps_1\inn N\inn \veps^S \inn N\inn \veps^A)^{\mu\nu}\nonumber\\
&&(\veps_1\inn N\inn \veps^A \inn V\inn \veps^S)^{\mu\nu}\quad , \quad(\veps_1\inn N\inn \veps^A \inn N\inn \veps^S)^{\mu\nu}\nonumber
\eeqa

where $\mu$ and $\nu$ are the world volume indices that contract with the volume $(p+1)$-form $\eps$ and $\mu\neq \nu$. We find the amplitude for one graviton and one B-field in terms of above RR structures:
\beqa
{\cal A} &\sim& \eps_{a_0\cdots a_3}\bigg[(p_2\inn N\inn \veps_1)^{a_3} \cM_1^{a_0a_1a_2}+(\veps_3^S \inn N\inn \veps_1)^{a_2a_3}\cM_2^{a_0a_1}+(p_2\inn V\inn \veps_3^S \inn N\inn \veps_1)^{a_3}\cM_3^{a_0a_1a_2}\nonumber\\
&&\quad\quad\quad+(p_2\inn N\inn \veps_3^S \inn N\inn \veps_1)^{a_3}\cM_4^{a_0a_1a_2}+(p_3\inn V\inn \veps_3^S \inn N\inn \veps_1)^{a_3}\cM_5^{a_0a_1a_2}\nonumber\\
&&\quad\quad\quad+(p_1\inn N\inn \veps_3^S \inn N\inn \veps_1)^{a_3}\cM_6^{a_0a_1a_2}+(\veps_1\inn N\inn \veps_3^S \inn V\inn \veps_2^A)^{a_2a_3}\cM_7^{a_0a_1}\nonumber\\
&&\quad\quad\quad+(\veps_1\inn N\inn \veps_3^S \inn N\inn \veps_2^A)^{a_2a_3}\cM_8^{a_0a_1}+(\veps_1\inn N\inn \veps_2^A \inn V\inn \veps_3^S)^{a_2a_3}\cM_9^{a_0a_1}\nonumber\\
&&\quad\quad\quad+(\veps_1\inn N\inn \veps_2^A \inn N\inn \veps_3^S)^{a_2a_3}\cM_{10}^{a_0a_1}+(p_3\inn N\inn \veps_2^A \inn N\inn \veps_1)^{a_0}\cM_{11}^{a_0a_1a_2}\nonumber\\
&&\quad\quad\quad+(p_3\inn V\inn \veps_2^A \inn N\inn \veps_1)^{a_0}\cM_{12}^{a_0a_1a_2}\bigg]+(2\leftrightarrow 3)\labell{Alast}
\eeqa
where
\beqa
{\cal M}_1^{a_0a_1a_2}&=&\frac{1}{4} \bigg[\bigg(p_2\inn V\inn\veps_3^S\inn N\inn p_1(\veps_2^A)^{a_1a_2}-2(p_3\inn V\inn\veps_2^A)^{a_2}(p_1\inn N\inn \veps_3^S)^{a_1}+(2\leftrightarrow 3)\bigg)p_2^{a_0}\cI_3 \nonumber\\
&&+\bigg(p_2\inn N\inn\veps_3^S\inn N\inn p_1(\veps_2^A)^{a_1a_2}-2(p_3\inn N\inn\veps_2^A)^{a_2}(p_1\inn N\inn \veps_3^S)^{a_1}+(2\leftrightarrow 3)\bigg)p_3^{a_0}\cI_2\nonumber\\
&&-\bigg(4p_3\inn V\inn\veps_3^S\inn N\inn p_1\cI_4-2p_1\inn N\inn\veps_3^S\inn N\inn p_1\cI_1\bigg)p_2^{a_0}(\veps_2^A)^{a_1a_2}\nonumber\\
&&+\bigg(8(p_3\inn V\inn\veps_3^A)^{a_2}\cI_4-2(p_1\inn N\inn\veps_3^A)^{a_2}\cI_1\bigg)p_2^{a_0}(p_1\inn N\inn \veps_2^S)^{a_1}\nonumber\\
&&-\bigg((p_1\inn N\inn\veps_3^S)^{a_0}(\veps_2^A)^{a_1a_2}+(2\leftrightarrow 3)\bigg) \bigg(2p_3\inn V\inn p_3\cI_4+p_2\inn V\inn p_3\cI_3-p_2\inn N\inn p_3\cI_2\bigg)\nonumber\\
&&-p_1\inn N\inn p_3\bigg(2p_3^{a_0}\cI_4(\veps_2^A)^{a_1a_2}\Tr[\veps_3^S\inn V]+ \cI_2(p_2\inn N\inn\veps_3^S)^{a_2}(\veps_2^A)^{a_0a_1}\nonumber\\
&&-\cI_3(p_2\inn V\inn\veps_3^S)^{a_2}(\veps_2^A)^{a_0a_1} -(2\leftrightarrow 3)-2 \cI_1(p_1\inn N\inn\veps_3^S)^{a_2}(\veps_2^A)^{a_0a_1}\nonumber\\
&&+2p_3^{a_0}\cG(\veps_2^A\inn V\inn\veps_3^S)^{a_1a_2}\cI_3+2p_3^{a_0}\cG(\veps_3^A\inn V\inn\veps_2^S)^{a_1a_2}\cI_2\bigg)\nonumber\\
&&-2p_2^{a_0}p_3^{a_1}\bigg(\cG(p_2\inn N\inn\veps_3^S\inn N\inn\veps_2^A)^{a_2}(\cJ_{16} -2\cJ_5)+2\cG(p_1\inn N\inn\veps_3^S\inn N\inn\veps_2^A)^{a_2}\cI_{2}\nonumber\\
&&+\cG(p_2\inn V\inn\veps_3^S\inn V\inn\veps_2^A)^{a_2}(\cJ_{16} -4\cJ+2\cJ_5)+4 \cG(p_3\inn V\inn\veps_3^S\inn V\inn\veps_2^A)^{a_2}\cJ_{12}\nonumber\\
&&+(\cG(p_2\inn V\inn\veps_3^S\inn N\inn\veps_2^A)^{a_2}+\cG(p_2\inn N\inn\veps_3^S\inn V\inn\veps_2^A)^{a_2})\cJ_{15}-2\cG(p_1\inn N\inn\veps_3^S\inn V\inn\veps_2^A)^{a_2}\cI_{3}\cr
&&-4\cG(p_3\inn V\inn\veps_3^S\inn N\inn\veps_2^A)^{a_2}\cJ_{4}-2p_3\inn V\inn p_3\cG(\veps_2^A\inn V\inn\veps_3^S)^{a_1a_2}\cJ_{12}\bigg)\cr
&&+p_1^{a_0}(\veps_2^A)^{a_1a_2}\bigg(p_2\inn N\inn\veps_3^S\inn N\inn p_2(\cJ_{16} -2\cJ_5)-4\cJ_{4} p_3\inn V\inn\veps_3^S\inn N\inn p_2\nonumber\\
&&+4\cI_2p_2\inn N\inn\veps_3^S\inn N\inn p_1+p_2\inn V\inn\veps_3^S\inn V\inn p_2(\cJ_{16}-4\cJ +2\cJ_5)+4p_3\inn V\inn\veps_3^S\inn V\inn p_2\cJ_{12}\nonumber\\
&&-2 p_2\inn V\inn\veps_3^S\inn N\inn p_1\cI_3+2p_2\inn V\inn\veps_3^S\inn N\inn p_2\cJ_{15}\nonumber\\
&&-2(2\cI_{4}p_1\inn N\inn p_3-\cJ_4p_2\inn N\inn p_3+\cJ_{12}p_2\inn V\inn p_3)\Tr[\veps_3^S\inn V]\bigg)\cr
&&+2 p_1^{a_0}(p_3\inn N\inn\veps_2^A)^{a_2}\bigg(2p_3^{a_1}\Tr[\veps_3^S\inn V]\cJ_{4}-2 \cI_{2}(p_1\inn N\inn\veps_3^S)^{a_1}-(p_2\inn N\inn\veps_3^S)^{a_1}\,(\cJ_{16} -2\cJ_5)\nonumber\\
&&(p_2\inn V\inn\veps_3^S)^{a_1}\cJ_{15}\bigg)-2p_1^{a_0}(p_3\inn V\inn\veps_2^A)^{a_2}\bigg(2p_3^{a_1} \Tr[\veps_3^S\inn V]\cJ_{12}+(p_2\inn N\inn\veps_3^S)^{a_1}\cJ_{15}\nonumber\\
&&+(p_2\inn V\inn\veps_3^S)^{a_1}\,(\cJ_{16}-4\cJ +2\cJ_5)\bigg)+4p_3\inn V\inn p_3  p_1^{a_0}\cG(\veps_3^A\inn V\inn\veps_2^S)^{a_1a_2}\cJ_4 +(2\leftrightarrow 3)\nonumber\\
&&-2p_2^{a_0}p_3^{a_1}\bigg(\cG(p_3\inn V\inn\veps_2^S\inn N\inn\veps_3^A)^{a_2}(\cJ_{16} +2\cJ_5)+\cG(p_3\inn N\inn\veps_2^S\inn V\inn\veps_3^A)^{a_2}\nonumber\\
&&\times(\cJ_{16}-4\cJ -2\cJ_5)+(\cG(p_3\inn V\inn\veps_2^S\inn V\inn\veps_3^A)^{a_2} +\cG (p_3\inn N\inn\veps_2^S\inn N\inn\veps_3^A)^{a_2})\cJ_{15}\bigg)\nonumber\\
&&-p_1^{a_0}(\veps_3^A)^{a_1a_2}\bigg(4(p_1\inn N\inn p_3\cI_{7}-p_3\inn V\inn p_3\cJ_{3})\Tr[\veps_2^S\inn V]+2p_3\inn V\inn\veps_2^S\inn N\inn p_1\, \cI_3\nonumber\\
&&+2p_3\inn V\inn\veps_2^S\inn N\inn p_3\,(\cJ_{16}-2\cJ )+(p_3\inn N\inn\veps_2^S\inn N\inn p_3\,+p_3\inn V\inn\veps_2^S\inn V\inn p_3)\cJ_{15}\bigg) \\
&&-2(p_2\inn V\inn\veps_3^A)^{a_2}\bigg(2p_2^{a_0}p_3^{a_1}\Tr[\veps_2^S\inn V] \cJ_{1}-2p_1^{a_0}(p_1\inn N\inn\veps_2^S)^{a_1}\cI_3-p_1^{a_0}(p_3\inn V\inn \veps_2^S)^{a_1}\cJ_{15}\nonumber\\
&& -p_1^{a_0}(p_3\inn N\inn\veps_2^S)^{a_1}(\cJ_{16} -4\cJ+2\cJ_5)\bigg)-4(p_1\inn N\inn\veps_3^A)^{a_2}\bigg(2p_2^{a_0}p_3^{a_1}\Tr[\veps_2^S\inn V] \cI_{7}\nonumber\\
&&-2 p_1^{a_0}(p_1\inn N\inn\veps_2^S)^{a_1}\cI_{1}+p_1^{a_0}(p_3\inn N\inn\veps_2^S)^{a_1}\cI_{3}-p_1^{a_0}(p_3\inn V\inn\veps_2^S)^{a_1}\cI_{2}\bigg)+8p_1^{a_0}(p_3\inn V\inn\veps_3^A)^{a_2}\nonumber\\
&&\times\bigg((p_3\inn N\inn\veps_2^S)^{a_1}\cJ_{12}-2 (p_1\inn N\inn\veps_2^S)^{a_1}\cI_{4}-(p_3\inn V\inn\veps_2^S)^{a_1}\cJ_{4}+2p_2^{a_1}\Tr[\veps_2^S\inn V] \cJ_{3}\bigg)\nonumber\\
&&+2p_1^{a_0}(p_2\inn N\inn\veps_3^A)^{a_2}\bigg( (p_3\inn N\inn\veps_2^S)^{a_1}\cJ_{15}+(p_3\inn V\inn\veps_2^S)^{a_1}(\cJ_{16} -2\cJ_5)+2p_2^{a_1}\Tr[\veps_2^S\inn V] \cJ_{2}\bigg)\nonumber\\
&&-4p_1\inn N\inn p_3p_1^{a_0} \bigg(\cG(\veps_3^A\inn V\inn\veps_2^S)^{a_1a_2}\cI_2+(2\leftrightarrow 3)\bigg) \bigg]\nonumber
\eeqa

\beqa
{\cal M}_2^{a_0a_1}&=&\frac{1}{4}\bigg[2(p_3\inn V\inn\veps_2^A)^{a_1}\bigg(p_1\inn N\inn p_2 p_2^{a_0}\cI_3+(2\leftrightarrow 3)\bigg)+2(p_3\inn N\inn\veps_2^A)^{a_1}\bigg(p_1\inn N\inn p_3 p_2^{a_0}\cI_3\nonumber\\
&&+(2\leftrightarrow 3)\bigg)-8(p_2\inn V\inn\veps_2^A)^{a_1}p_1\inn N\inn p_3 p_3^{a_0}\cI_7+4(p_1\inn N\inn\veps_2^A)^{a_1}p_1\inn N\inn p_3 p_3^{a_0}\cI_1\nonumber\\
&&-(\veps_2^A)^{a_0a_1}\bigg(2p_1\inn N\inn p_2 p_3\inn V\inn p_3\cI_4+p_1\inn N\inn p_3 
(p_2\inn V\inn p_3\cI_2-p_2\inn N\inn p_3\cI_3)\bigg)\nonumber\\
&&-p_2^{a_0}\bigg(p_3^{a_1}(p_3\inn V\inn\veps_2^A\inn V\inn p_2\cJ_1-\frac{1}{2}p_3\inn V\inn\veps_2^A\inn N\inn p_1\cI_3\nonumber\\
&&+p_2\inn V\inn\veps_2^A\inn N\inn p_3\cJ_2 -p_3\inn V\inn\veps_2^A\inn N\inn p_3(\cJ+\cJ_5)+\frac{1}{2}p_3\inn N\inn\veps_2^A\inn N\inn p_1\cI_2)\nonumber\\
&&+\frac{1}{2}(p_3\inn V\inn\veps_2^A)^{a_1}\left(p_2\inn V\inn p_2\cJ_1+p_3\inn V\inn p_3\cJ_4-2p_2\inn N\inn p_3\cJ\right)-2(p_2\inn V\inn\veps_2^A)^{a_1}p_3\inn V\inn p_3\cJ_3\nonumber\\
&&+\frac{1}{2}(p_3\inn N\inn\veps_2^A)^{a_1}\left(p_2\inn N\inn p_3\cJ_{15}+p_2\inn V\inn p_3(\cJ_{16} -2\cJ)\right)+(p_1\inn N\inn\veps_2^A)^{a_1}p_3\inn V\inn p_3\cI_4 \bigg)\bigg]\nonumber
\eeqa

\beqa
{\cal M}_3^{a_0a_1a_2}&=&-\frac{1}{4}\bigg[(\veps_2^A)^{a_1a_2} \bigg(p_1\inn N\inn p_2 p_2^{a_0}\cI_3+(2\leftrightarrow 3)\bigg)+p_2^{a_0}p_3^{a_1}\bigg(\frac{1}{2}(p_1\inn N\inn\veps_2^A)^{a_2}\cI_3\nonumber\\
&&+(p_2\inn V\inn\veps_2^A)^{a_2}\cJ_1 -(p_3\inn N\inn\veps_2^A)^{a_2}\cJ_5\bigg)\nonumber\\
&&+\frac{1}{4}p_1^{a_0}(\veps_2^A)^{a_1a_2}\bigg(p_2\inn V\inn p_2\cJ_1-p_3\inn V\inn p_3\cJ_4+2p_2\inn N\inn p_3\cJ\bigg)\bigg]\nonumber
\eeqa

\beqa
{\cal M}_4^{a_0a_1a_2}&=&-\frac{1}{4}\bigg[(\veps_2^A)^{a_1a_2} \bigg(p_1\inn N\inn p_3 p_2^{a_0}\cI_3+(2\leftrightarrow 3)\bigg)-p_2^{a_0}p_3^{a_1}\bigg(\frac{1}{2}(p_1\inn N\inn\veps_2^A)^{a_2}\cI_2\nonumber\\
&&+(p_3\inn V\inn\veps_2^A)^{a_2}\cJ_5 -(p_2\inn V\inn\veps_2^A)^{a_2}\cJ_2\bigg)\nonumber\\
&&+\frac{1}{4}p_1^{a_0}(\veps_2^A)^{a_1a_2}\bigg(p_2\inn N\inn p_3\cJ_{15}-p_2\inn V\inn p_3(\cJ_{16} -2\cJ)\bigg)\bigg]\nonumber
\eeqa

\beqa
{\cal M}_5^{a_0a_1a_2}&=&\frac{1}{4}\bigg[4(\veps_2^A)^{a_1a_2} p_1\inn N\inn p_2 p_2^{a_0}\cI_4 +p_2^{a_0}p_3^{a_1}\bigg(2(p_1\inn N\inn\veps_2^A)^{a_2}\cI_4 -4 (p_2\inn V\inn\veps_2^A)^{a_2}\cJ_3\nonumber\\
&& -(p_3\inn N\inn\veps_2^A)^{a_2}\cJ_{12}+(p_3\inn V\inn\veps_2^A)^{a_2}\cJ_4\bigg)+p_1^{a_0}(\veps_2^A)^{a_1a_2}p_2\inn V\inn p_2\cJ_3\bigg]\nonumber
\eeqa

\beqa
{\cal M}_6^{a_0a_1a_2}&=&-\frac{1}{2}\bigg[(\veps_2^A)^{a_1a_2} p_1\inn N\inn p_2 p_2^{a_0}\cI_1+2 p_2^{a_0}p_3^{a_1}\bigg((p_1\inn N\inn\veps_2^A)^{a_2}\cI_1-\frac{1}{2}(p_3\inn V\inn\veps_2^A)^{a_2}\cI_2\nonumber\\
&&+\frac{1}{2}(p_3\inn N\inn\veps_2^A)^{a_2}\cI_3 +2 (p_2\inn V\inn\veps_2^A)^{a_2}\cI_7\bigg)+p_1^{a_0}(\veps_2^A)^{a_1a_2}p_2\inn V\inn p_2\cI_7\bigg]\nonumber
\eeqa

\beqa
{\cal M}_7^{a_0a_1}={\cal M}_8^{a_0a_1}= p_2^{a_0}p_3^{a_1}p_2\inn N\inn p_3\cJ\nonumber
\eeqa
\beqa
{\cal M}_9^{a_0a_1}={\cal M}_{10}^{a_0a_1}=p_2^{a_0}p_3^{a_1}p_2\inn V\inn p_3\cJ\nonumber
\eeqa
\beqa
{\cal M}_{11}^{a_0a_1a_2}=p_2^{a_0}p_3^{a_1} (p_2\inn N\inn\veps_3^S)^{a_2}\cJ\nonumber
\eeqa
\beqa
{\cal M}_{12}^{a_0a_1a_2}=p_2^{a_0}p_3^{a_1} (p_2\inn V\inn\veps_3^S)^{a_2}\cJ\nonumber
\eeqa

where  $\cJ,\cJ_1,\cJ_2,\cJ_3,\cJ_4,\cJ_5,\cJ_{12},\cJ_{15}$ and $\cJ_{16}$ are  new integrals which appear in this case. The explicit form of these integrals has been
found in \cite{Garousi:2011bm,Komeil:2013prd}. The operator $\cal G$, which conspicuouses in ${\cal M}_{1}^{a_0a_1a_2}$, is defined as \cite{Komeil:2013prd}:
 \beqa
{\cal G} (\veps_n^A\inn V\inn\veps_m^S)^{\mu \nu}&\rightarrow &(\veps_n^A\inn V\inn\veps_m^S)^{\mu \nu}-(\veps_n^S\inn N\inn\veps_m^A)^{\mu \nu}\cr
{\cal G} (\veps_n^S\inn V\inn\veps_m^A)^{\mu \nu}&\rightarrow &(\veps_n^S\inn V\inn\veps_m^A)^{\mu \nu}-(\veps_n^A\inn N\inn\veps_m^S)^{\mu \nu}\cr
{\cal G} (\veps_n^A\inn N\inn\veps_m^S)^{\mu \nu}&\rightarrow &(\veps_n^A\inn N\inn\veps_m^S)^{\mu \nu}-(\veps_n^S\inn V\inn\veps_m^A)^{\mu \nu}\cr
{\cal G} (\veps_n^S\inn N\inn\veps_m^A)^{\mu \nu}&\rightarrow &(\veps_n^S\inn N\inn\veps_m^A)^{\mu \nu}-(\veps_n^A\inn V\inn\veps_m^S)^{\mu \nu}\cr
{\cal G} (p\inn \veps_n^A\inn V \inn \veps_m^S)^{\mu}&\rightarrow &(p\inn \veps_n^A\inn V \inn \veps_m^S)^{\mu}-(p\inn \veps_n^S\inn N \inn \veps_m^A)^{\mu} 
\eeqa
 where $n$ and $m$ are the particle labels of the polarization tensors.

 Using the $SL(2,R)$ symmetry fixing in \cite{Becker:2010ij}, one finds the low energy expansion of the integrals $\cJ,\cJ_1,\cJ_2,\cJ_3,\cJ_4,\cJ_5,\cJ_{12},\cJ_{15}$ and $\cJ_{16}$, for the general setup. Considering the expansion of integrals $\cI_1,\cI_2,\cI_3,\cI_4,\cI_{7}$ in the sections 2.2.1 and 2.2.2, it is clear that only the integrals $\cI_4,\cI_7, \cJ_1, \cJ_4, \cJ_3$ have massless open pole as well as closed pole. The other integrals have only closed pole. Using the open pole terms in the low energy expansion of these integrals, one can find the string open pole amplitude at low energy.

We find the exact consistency between this amplitude and corresponding field theory amplitude. The amplitude \reef{Alast} satisfies the NSNS ward identity; however, it dose not satisfy the RR ward identity. It can easily be extended to the RR invariant amplitude by including the amplitude of the RR $(p-1)$-form with three, two and no transvers indices, which the first two cases has been found in sections 2.2.1 and 2.2.2. In this work, we are not interested in the case that the RR potential carry no transvers index. We then check that this amplitude is exactly equal to the result of T-dual ward identity in \cite{Komeil:2013prd}.

\end{document}